\documentstyle[12pt]{article}
\date{}

\newskip\humongous \humongous=0pt plus 1000pt minus 1000pt
  \newif\ifdtup

\def\abar{{\bar \alpha_S}}
\def\abn{{{\bar \alpha_S} \over N}}
\def\as{\alpha_S}
\def\ltap{\raisebox{-.55ex}{\rlap{$\sim$}} \raisebox{.55ex}{$<$}}
\def\gtap{\raisebox{-.55ex}{\rlap{$\sim$}} \raisebox{.55ex}{$>$}}

\def\kper{k_{\perp}}
\def\cf{{\cal F}}
\def\ga{\gamma}

\def\bk{{\mbox{\bf k}}}
\def\bq{{\mbox{\bf q}}}

\def\cav#1{Cambridge preprint Cavendish--HEP--#1}

\def\np#1#2#3{Nucl.\ Phys.\ B#1 (19#3) #2}
\def\pl#1#2#3{Phys.\ Lett.\ #1B (19#3) #2}
\def\pr#1#2#3{Phys.\ Rev.\ D #1 (19#3) #2}
\def\prep#1#2#3{Phys.\ Rep.\ #1 (19#3) #2}

\def\zp#1#2#3{Zeit.\ Phys.\ C#1 (19#3) #2}

\begin{document}
\renewcommand{\thefootnote}{\fnsymbol{footnote}}

\begin{titlepage}

\thispagestyle{empty}

\begin{flushright}
     DFF 207/6/94\\   June 1994
     \end{flushright}
\vspace*{5mm}

\begin{center}{\large LOW-$x$ STRUCTURE FUNCTIONS\footnote{Invited
talk at Les Recontres de Physique de La Vallee d'Aoste, {\it
Results and Perspective in Particle Physics}, La Thuile, March 1994 }}
\end{center}

\vskip 1 cm

\begin{center}
{\bf Stefano Catani}\\

\vskip 5 mm

{I.N.F.N., Sezione di Firenze}\\
{and Dipartimento
di Fisica, Universit\`a di Firenze}\\
{Largo E. Fermi 2, I-50125 Florence, Italy}
\end{center}

\vskip 2 cm

\begin{abstract}
QCD predictions on the low-$x$ behaviour of the structure functions are
reviewed and compared with the recent measurements of $F_2(x,Q^2)$ at HERA.
The present theoretical accuracy of these predictions is discussed.

\end{abstract}

\end{titlepage}
%\newpage

\section{Introduction}
\label{intro}

At present and future hadron colliders, precise quantitative
tests of QCD and searches for new physics within the standard
model (top and Higgs production) and beyond are carried out in the
small-$x$ kinematic region. By small $x$ we mean that the ratio
$x= Q^{2}/S$, between the typical transferred momentum $<p_{t}>
= Q$ in the process and the centre-of-mass energy $\sqrt{S}$ of
the colliding hadrons, is much smaller than unity. For these
small-$x$ processes reliable and accurate theoretical predictions
are clearly necessary.

Despite the general relevance of this topic, the present
contribution does not intend to be a comprehensive (even if concise)
review on the theory of small-$x$ physics. I shall concentrate
on the low-$x$ behaviour of the (proton) structure functions in
deep inelastic electron-proton scattering. There are two very
simple motivations for limiting ourselves to considering structure
functions. First, the study is relevant by itself for
investigating hadron physics at very high energy (since
$S=Q^{2}/x$, the small-$x$ limit at fixed $Q^{2}$ is equivalent to
$S\rightarrow \infty$). Second, according to the parton model the
structure functions are proportional to the parton densities of
the hadron and their knowledge is essential for
accurate phenomenology at {\em any} hadron collider.

In the following, I shall review some of the main QCD predictions
about the small-$x$ behaviour of the structure functions and, in
particular, I shall try to discuss the {\em present accuracy} of
these predictions. Some other aspects of small-$x$ physics are
considered elsewhere at this meeting [\ref{Pope}].

\section{Structure functions at HERA}
\label{sf at H}

The natural starting point of our discussion on low-$x$ structure
functions is represented by the recent HERA data on the proton
structure function $F_{2}$ [\ref{H1},\ref{ZEUS},\ref{Newton}].

HERA is the first electron-proton collider ever built and is
operating at present with the centre-of-mass energy $\sqrt{S} =
296$ GeV. Due to this large centre-of-mass energy, it is able to
see into very small values of the Bjorken variable $x, \; x \, \gtap
\,10^{-4}$, even in the deep-inelastic-scattering (DIS)
region 10 GeV${}^{2} \,\ltap \,Q^{2} \,\ltap \,10^{5}$ GeV${}^{2}$ (I
use standard notations for the DIS kinematic variables, namely
$Q^{2} = -q^{2}, \;x =Q^{2}/2p \cdot q, \;y=p \cdot q/p\cdot k$ with
$S=Q^{2}/xy$; see Fig.~1) This is an entirely new DIS regime which
extends for two orders of magnitude, both in $x$ and $Q^{2}$, outside
that ($x \,\gtap \,10^{-2}, \;Q^{2} \,\ltap \,10^3$ GeV${}^{2}$)
investigated by previous fixed-target experiments [\ref{Aachen}].

For values of $Q^{2} \ltap 10^3$ GeV${}^2$, the electron-proton
inclusive cross
section is dominated by the exchange of a single off-shell photon
and given by
\begin{equation}
\frac{d \sigma^{ep}}{dx \; dQ^{2}} = \frac{4\pi \alpha^{2}}{xQ^{4}}
\left( 1 -y +\frac{1}{2} \; \frac{y^{2}}{1+R(x,Q^{2})} \right) F_{2}
(x,Q^{2}) \; .
\label{dsigma}
\end{equation}
Here $F_{2}$ is the customary proton structure function, which is
proportional to the sum $\sigma _{L} + \sigma_{T}, \; \sigma_{i}$'s
being the cross sections off a transversely $(i=T)$ or
longitudinally $(i=L)$ polarized photon. The function $R$ denotes
the ratio $R=\sigma _{L}/\sigma_{T}$: it is smaller than 10\%
in the region which is being studied at HERA. It follows that
the HERA data on the cross section (\ref{dsigma}) provides a
direct measurement of $F_{2}(x,Q^{2})$.

The data collected at HERA in 1992 [\ref{H1},\ref{ZEUS}] have already
shown clearly that $F_{2}(x,Q^{2})$ increases very steeply at
small $x$. This behaviour has to be contrasted with that measured
at higher $x$-values (Fig.~2). Previous data in the region of $x
\sim 10^{-2}$ are indeed consistent with a constant (or almost
constant) behaviour of $F_{2}$ for $x \rightarrow 0$. Moreover,
HERA data on $F_{2}$ as to be regarded as a qualitatively new
(although, possibly, not unexpected) experimental result: a
similar strong rise with the energy has not been observed so far in
total hadronic cross sections.

\vspace*{7.5 cm}

\noindent {\bf Fig. 1}: Kinematic variables for deep inelastic electron-proton
scattering.

\vspace*{14 cm}

\noindent {\bf Fig. 2}: H1 data on $F_2(x,Q^2)$. Data points of the NMC and
BCDMS experiments are shown for comparison.

\section{Total hadronic cross sections}
\label{THCS}

A compilation of data [\ref{PDG}] on total
hadronic cross sections is reported in Fig.~3.
We can see that they increase slowly with
increasing $\sqrt{S}$.

\vspace*{16.5 cm}

\noindent {\bf Fig. 3}: Data for total cross sections with fits of type
(\ref{sigtot}).

\vspace*{1 cm}

The curves in Fig.~3 are the results of a Regge-type fit performed
by Donnachie and Landshoff [\ref{DOLA}]. It provides a very simple
and economical description of all data in the form:
\begin{equation}
\sigma_{TOT} = X \; S^{\varepsilon} + Y \; S^{-\eta} \;.
\label{sigtot}
\end{equation}
The parameters $X$ and $Y$ depend on the type of colliding hadrons
whilst the powers $\varepsilon$ and $\eta$ are constrained to be
`universal' and found (by the fit) to be $\varepsilon = 0.08$ and
$\eta = 0.45$. In this Regge-type phenomenological model
$\alpha_{P}(0) =1+\varepsilon$ and $1-\eta$ are thus respectively
interpreted as the intercept of pomeron trajectory (it carries
the vacuum quantum numbers) and of the $\rho$-meson trajectory
$(\rho , \omega , f,a,)$.

The increase of the cross section (\ref{sigtot}) with the energy
is due to the pomeron exchange and is indeed very slowly because
the pomeron trajectory is above unity be a small amount, i.e.
$\varepsilon = 0.08$. Actually, since $\varepsilon \ll 1$, to a good
approximation we have $\; S^{\varepsilon} \simeq 1 + \varepsilon \ln S
+ \frac{1}{2} \varepsilon^{2}  \ln^{2}S$. This is the reason why
double-logarithmic expressions of the type $\: \sigma_{TOT} = A
S^{-n} + B + C \ln S + D \ln^{2} S$ [\ref{PDG}] are equally
successful in describing the data.

It is worth noting that this slow (approximately logarithmic)
increase of the cross section holds true also for the
photoproduction case. HERA data on the $\gamma p$ cross section
[\ref{phot}] (Fig.~4)
are in remarkable agreement with the value predicted by Donnachie
and Landshoff almost ten years ago [\ref{DOLA1}].

\vspace*{8.5 cm}

\noindent {\bf Fig. 4}: Fit of type (\ref{sigtot}) for the $\gamma p$ cross
section data below ${\sqrt S}=20$ GeV. The two data points at ${\sqrt S}= 200$
GeV correspond to the H1 and ZEUS measurements at HERA.

\vspace*{1 cm}

On the basis of this successful phenomenological description of the
energy dependence of total cross sections, one can assume
a corresponding $x$-dependence for the structure function $F_{2}$
[\ref{ALLM},\ref{DOLA2}]. In particular, one can extrapolate the
Regge-type parametrization (\ref{sigtot}) from the photoproduction
limit $(Q^{2} = 0)$ to the low-$Q^{2}$ $(Q^{2} \,\ltap$ 10
GeV${}^{2}$) regime by using the following expression [\ref{DOLA2}]
\begin{equation}
F_{2}(x,Q^{2}) = A \; x^{-0.08} \;\frac{Q^{2}}{Q^{2}+a^{2}} + B \;
x^{0.45} \; \frac{Q^{2}}{Q^{2}+b^{2}}\; .
\label{F2x}
\end{equation}
This extrapolation is obtained by a simple dipole model $(a \simeq
$ 750 MeV) for the $Q^{2}$-dependence: $F_{2}$ is forced to vanish
for $Q^{2} \rightarrow 0$ (gauge invariance) and to be $Q^2$-independent
at high $Q^2$ (Bjorken scaling).

The comparison between the parametrization (\ref{F2x}) and HERA
data (Fig.~5) shows that this type of models, based on the energy
behaviour of low-$p_{t}$ hadronic interaction (soft processes),
fails in describing the structure function $F_{2}(x,Q^{2})$ as
measured in the DIS regime. Hard physics is necessary for
explaining the $x$-shape: a steeper $F_{2}$ is indeed expected from
hard (perturbative) QCD.

\pagebreak

\vspace*{18 cm}

\noindent {\bf Fig. 5}: Comparison of the measured  $F_2(x,Q^2)$ (H1 data)
with the
parametrization in Eq.~(\ref{F2x}) (DOLA) and with several QCD expectations
(see Sect.~5). The overall normalization uncertainty of $8\%$ is not included
in the error bars.

%\vspace*{1 cm}

\section{Structure functions in perturbative QCD}
\label{SFPQ}

\subsection{The QCD improved parton model}

According to the na\"{\i}ve parton model the hadron is built up by
point-like partons (quarks and gluons) which are almost
non-interacting and scattered incoherently when their transverse
momentum $p_{\perp i}$ is larger than some value $Q_{0}$ of the order of
the inverse hadronic radius $1/R_{h} \sim 1$~GeV. Therefore, in
the large-$Q^{2}$ limit $(Q^{2} \gg Q^{2}_{0})$ the structure
function $F_{2}(x,Q^{2})$ is given by
\begin{equation}
F_{2}(x,Q^{2}) = \sum^{N_{f}}_{i=1} e^{2}_{i} \left[f_{q_i} (x)
+  f_{\bar q_i} (x)\right] \; ,
\label{F2(}
\end{equation}
where $f_{q_i} (x) \equiv x q_i(x)$ are the {\em scale-invariant}
quark densities of the hadron and $e_{i}$ are the quark
charges.

Perturbative QCD modifies this simple picture. The
large $k_{\perp}$-gap $(Q^{2}_{0} \ll k_{\perp} \ll Q^{2}_{0})$
between the incoming proton and the scattered quark can be filled
up by QCD radiation of massless partons. They have a logarithmic
spectrum $d k_{\perp}^{2}/k_{\perp}^{2}$ in transverse momentum
thus leading to corrections of the type
\begin{equation}
\alpha_{S} \int^{Q^{2}}_{Q^{2}_{0}} \frac{d
k_{\perp}^{2}}{k_{\perp}^{2}} = \alpha_{S} \ln \frac{Q^{2}}{Q^{2}_{0}}
\label{alphaS}
\end{equation}
that can be large $(\alpha_{S}  \ln Q^{2}/Q^{2}_{0} \sim 1$ when
$Q^{2} \gg Q^{2}_0)$ even if the partons are weakly coupled
$(\alpha_{S} (Q^{2}) \ll 1)$. The large logarithmic contributions
in Eq.~(\ref{alphaS}) have to be computed to all orders in
$\alpha_{S}$ and their resummation gives rise to parton densities
evolving with $Q^{2}$ (scaling violations) in a predictable way.

The QCD improved parton model is summarized by the following
factorization formula [\ref{Alt}]
\begin{equation}
F_{2}(x,Q^{2}) = \sum_{Q=q,{\bar q}, g} \int^{1}_{x} \frac{dz}{z}
\; {\hat F}_{2a}(x/z, Q^{2}; Q^{2}_{0})
\,f_{a}(z,Q^{2}_{0}) + {\cal O}\left( 1/Q^{2}\right) \; .
\label{F2(x}
\end{equation}
Here $f_{a} (x,Q^{2}_{0})$ are the input (non-perturbative)
parton densities $(f_{g}(x) \equiv xg(x)$ is the gluon density) at
the scale $Q^{2}_{0}, \; {\hat F}_{2a}$ accounts for the QCD
evolution from $Q^{2}_{0}$ to $Q^{2}$ and is computable in
perturbation theory and the ${\cal O}(1/Q^{2})$ term stands for the
so-called higher-twist contributions, i.e. contributions vanishing
as inverse powers of $Q^{2}$ in the high-$Q^{2}$ regime.

Note that the $Q^{2}$-dependence is completely factorized in
Eq.~(\ref{F2(x}) and thus predicted by the perturbative QCD
component ${\hat F}_{2a}$. On the contrary, the $x$-dependence of
$F_{2}(x,Q^{2})$ enters Eq.~(\ref{F2(x}) through the convolution
of ${\hat F}_{2a}$ and $f_{a}$ with respect to the
longitudinal-momentum fraction $z$. As a result, the $x$-shape of
$F_{2}$ is determined both by the input parton density and by its
perturbative evolution. For instance, if we consider a model in
which both the input and the evolution have a power dependence on
$x$, i.e. $f_{a}(x) \sim x^{-\lambda}$ and ${\hat F}_{2a}(x) \sim
x^{-\alpha}$,  from Eq.~(\ref{F2(x}) we see that the steeper one
wins as $x\rightarrow 0$ : $F_{2}(x) \sim x^{-\max \{ \alpha ,
\lambda\} }$.

The perturbative QCD component ${\hat F}_{2a}$ in Eq.~(\ref{F2(x})
is given by
\begin{equation}
{\hat F}_{2a} (x,Q^{2};Q^{2}_{0}) = \sum_{b} \int^{1}_{x}
\frac{dz}{z} \; C_{2b}(x/z, \alpha_{S}(Q^{2}) )
\;F_{ba}(z,Q^{2};Q^{2}_{0})\; ,
\label{hatF}
\end{equation}
where $C_{2a}$ is the process-dependent coefficient function and
$F_{ab}$ , the universal (i.e. process independent) Green function
for the evolution of the parton densities, fulfils the following
generalized Altarelli-Parisi equation
\begin{equation}
F_{ab}(x,Q^{2};Q^{2}_{0}) = \delta_{ab} \;\delta(1-x) + \sum_{c}
\int^{Q^{2}}_{Q^{2}_{0}} \frac{dk_{\perp}^{2}}{k_{\perp}^{2}}
\int^{1}_{x} dz \; P_{ac} (z,\alpha_{S}(k_{\perp}^{2}))
\; F_{cb}(x/z, k_{\perp}^{2}; Q^{2}_{0} )\; .
\label{Fab}
\end{equation}
The parton densities at the hard scale $Q^{2}$ are obtained from the
%given in terms of
the corresponding inputs as follows
\begin{equation}
f_{a} (x,Q^{2}) = \sum_{b} \int^{1}_{x}
\frac{dz}{z}
\; F_{ab}(x/z,Q^{2};Q^{2}_{0}) \;f_{b}(z,Q^{2}_{0})\; .
\label{fax}
\end{equation}

The coefficient function $C_{2}(z,\alpha_{S})$ and the
Altarelli-Parisi splitting function $P(z,\alpha_{S})$ are both
computable in QCD perturbation theory and their power series
expansion has the form:
\begin{equation}
C_{2}(z, \alpha_{S}) = \delta(1-z) + \alpha_{S} \;C_{2}^{(1)}(z)
+ \alpha_{S}^{2} \; C_{2}^{(2)}(z) + \dots \; ,
\label{C2}
\end{equation}
\begin{equation}
P(z, \as) = \alpha_{S} \; P^{(1)}(z) + \alpha_{S}^{2}
\; P^{(2)}(z) + \alpha_{S}^{3} \; P^{(3)}(z) + \dots \; ,
\label{Pz}
\end{equation}
The coefficient function $C_{2}$ measures the effective coupling
$\gamma^{*}$-parton whereas Eq.~(\ref{Fab}) has a simple
partonic interpretation. It describes a generalized parton
evolution along a space-like cascade in which the successive
$k_{\perp}$'s are strongly ordered (Fig.~6). The Altarelli-Parisi
splitting function $P_{ab}(z, \alpha_{S})$ can be consistently
interpreted as the probability of emission of a bunch of partons
(jets) whose transverse momenta are of the same order. In
particular $P^{(1)}, P^{(2)}, \dots , P^{(n)}$, in Eq.~(\ref{Pz})
respectively represent the emission of one, two, ... , $n$
partons with almost equal transverse momenta (Fig.~6b). We see that
the radiation of an additional parton without $k_{\perp}$-ordering
costs an extra power of $\alpha_{S}$ and, hence, is a subdominant
effect with respect to the $Q^{2}$-evolution of $F_{2}(x,Q^{2})$.

\vspace*{10 cm}

\noindent {\bf Fig. 6}: Partonic picture of the perturbative QCD evolution:
$(a)$ strong ordering of transverse momenta
$(Q^2>k_{\perp n}^2> \dots >k_{\perp 1}^2>Q_0^2)$ along the space-like cascade;
$(b)$ splitting function for the emission of partons with comparable transverse
momenta.

\vspace*{1 cm}

For the sake of simplicity, in the following I do not consider the higher-order
contributions $C_2^{(n)} \;(n \geq 1)$ to the coefficient function. Note,
however, that by no means this implies that their quantitative effect is
negligible in the small-$x$ region [\ref{CCH},\ref{Neerven},\ref{CH}].

In order to proceed in our discussion, it is also convenient to
introduce $N$-moments as follows
\begin{equation}
F_{N}(Q^{2}; Q^{2}_{0}) \equiv \int^{1}_{0} dx \; x^{N-1} \;F(x,Q^{2};
Q^{2}_{0}) \; .
\label{FN}
\end{equation}
The limit $x \rightarrow 0$ corresponds to the small-$N$ limit in
the $N$-moment space.

The Mellin transformation in Eq.~(\ref{FN})
diagonalizes the convolution over $z$ in Eq.~(\ref{Fab}). The
solution of the latter is
\begin{equation}
F_{N}(Q^{2}; Q^{2}_{0}) = \exp \left\{ \int^{Q^{2}}_{Q^{2}_{0}}
\frac{dk_{\perp}^{2}}{k_{\perp}^{2}}
\;\gamma_{N}(\alpha_{S}(k_{\perp}^{2}))\right\}\; ,
\label{FNQ}
\end{equation}
where the anomalous dimensions $\gamma_{N}$ are related to the
$N$-moments of the splitting functions as follows
\begin{equation}
\gamma_{N}(\alpha_{S}) = \int^{1}_{0} dz \;z^{N} \;P(z,\alpha_{S}) =
\alpha_{S} \;\gamma_{N}^{(1)} + \alpha_{S}^{2} \;\gamma_{N}^{(2)} +
\dots \;\;\; .
\label{gammaN}
\end{equation}

The expression (\ref{FNQ}) explicitly resums the large logarithmic
contributions $\alpha_{S}^{n} (\ln Q^{2}/Q^{2}_{0})^{k}$ we have
mentioned before. Inserting the one-loop $(\gamma^{(1)}_{N})$,
two-loop $(\gamma^{(2)}_{N})$, ... anomalous dimensions into
Eq.~(\ref{FNQ}), one can systematically resums leading $(k=n)$,
next-to-leading $(k=n-1)$, ... logarithmic terms.

\subsection{Altarelli-Parisi evolution to leading order}

The perturbative QCD evolution of the parton densities to leading
order is particularly simple in the small-$x$ limit [\ref{Alt}]. In
this case the gluon channel dominates because the splitting function
\begin{equation}
P_{gg}(z,\alpha_{S}) \simeq \frac{C_{A}\alpha_{S}}{\pi}
\,\frac{1}{z} + {\cal O}(\alpha_{S}^{2}), \;\;\;\;\; \;\;(z\rightarrow 0) \; ,
\label{Pgg}
\end{equation}
has a $1/z$ singularity, related to the exchange of a spin-one
particle, the gluon, in the $t$-channel. The corresponding
anomalous dimension $({\bar \alpha}_{S} \equiv C_{A}
\alpha_{S}/\pi)$
\begin{equation}
\gamma_{gg,N} (\alpha_{S}) \simeq \frac{{\bar \alpha}_{S}}{N} +
{\cal O}(\alpha_{S}^{2}) \;\;,
\; \;\;\;(N\rightarrow 0) \;,
\label{gammagg}
\end{equation}
is singular for $N\rightarrow 0$. Inserting Eq.~(\ref{gammagg})
into Eq.~(\ref{FNQ}) and transforming back to the $x$-space, one
finds:
%\footnote{For the sake of simplicity, we consider the case of
%a fixed coupling $\alpha_{S}$.}:
\begin{equation}
F_{gg} (x,Q^{2};Q^{2}_{0}) \simeq \exp \left\{ 2\; \sqrt{{\bar
\alpha}_{S} \ln \frac{1}{x} \,\ln \frac{Q^{2}}{Q^{2}_{0}}} \right\}
\; .
\label{Fgg}
\end{equation}

The result in Eq.~(\ref{Fgg}) shows that, due to the
perturbative QCD evolution, the parton densities raise rapidly for
$x\rightarrow 0$. Their increase is faster than any power of $\ln
x$, although slower than any power of $1/x$. Note also that,
because of the factor $\ln Q^{2}/Q^{2}_{0}$, the effective slope in
$x$ increases proportionally to the evolution range in
$k_{\perp}^{2}$.

Equation (\ref{Fgg}) resums the double logarithmic contributions
$(\alpha_{S} \ln \frac{1}{x} \ln \frac{Q^{2}}{Q^{2}_{0}})^{n}$ to
all orders in $\alpha_{S}$. It has been obtained by assuming a
fixed coupling $\alpha_{S}$. The inclusion of the running coupling
and of the next-to-leading term $\gamma^{(2)}_{N}$ for the
anomalous dimensions does not change qualitatively the small-$x$
behaviour of $F(x,Q^{2};Q^{2}_{0})$ (see Sect.~5).

In order to gain physical insight into the result (\ref{Fgg}), let us consider
a simplified derivation. The probability of emission of a gluon with a large
rapidity $y=\ln 1/z$ and transverse momentum $\kper$ is:
\begin{equation}
\label{spec}
dw = \abar \;dy \;\frac{d\kper^2}{\kper^2} \;\;.
\end{equation}
The Green function $F_{gg}$ is obtained by integrating the spectrum
(\ref{spec}) over the rapidity range $L=\ln 1/x$ and summing over any number
$n$ of final state gluons as follows
\begin{eqnarray}
\label{fsum}
F_{gg}(x, Q^2; Q_0^2) &\sim& \sum_n \frac{1}{n!} \abar^n \int_0^L dy_1 \dots
\int_0^L dy_n \int \frac{dk_{\perp 1}^2}{k_{\perp 1}^2} \dots
\frac{dk_{\perp n}^2}{k_{\perp n}^2} \;
\Theta(Q^2>k_{\perp n}^2> \dots >k_{\perp 1}^2>Q_0^2)
\nonumber \\
&=&\sum_n \frac{1}{n!} \,(\abar L)^n \;
\int \frac{dk_{\perp 1}^2}{k_{\perp 1}^2} \dots
\frac{dk_{\perp n}^2}{k_{\perp n}^2}
\; \Theta(Q^2>k_{\perp n}^2> \dots >k_{\perp 1}^2>Q_0^2)
\;\;.
\end{eqnarray}
The statistical factor $1/n!$ in Eq.~(\ref{fsum}) is due to the
identity of the $n$ final-state gluons. The $\Theta$-function denotes the
constraint of transverse-momentum ordering which is appropriate for the
$Q^2$-evolution to leading accuracy. After $\kper$-integration, this constraint
produces an additional factor of $1/n!$, thus leading to $(\, (n!)^2 \simeq
(2n)!/2^{2n} \;{\mbox {\rm for}}\; n \gg 1)$
\begin{equation}
\label{fgap}
F_{gg} (x,Q^{2};Q^{2}_{0}) \simeq \sum_n \left(\frac{1}{n!}\right)^2
\;\left( \sqrt{{\bar \alpha}_{S} L \;\ln (Q^2/Q^2_0)}\right)^{2n}
\simeq \exp \left\{ 2 \; \sqrt{{\bar
\alpha}_{S} L \;\ln (Q^{2}/Q^{2}_{0})} \right\} \; .
\end{equation}

This discussion shows that the rise of the parton densities at small $x$
is due to the perturbative emission of many gluons (each of them giving a
constant amplitude) over a large rapidity gap. Actually, at very small values
of $x$, the $L$ factors associated with the large rapidity gap may dominate
the analogous factors $\ln Q^2/Q_0^2$ due to the $\kper$-evolution. In this
case one has to go beyond the leading logarithmic approximation in
Eqs.~(\ref{Pgg})-(\ref{Fgg}). The higher-order terms $P^{(n)}(z) \;(n>1)$ in
the splitting functions (\ref{Pz}) can be large and have to be taken into
account. In fact, the additional powers of $\as$ due to the emission of gluons
with comparable $\kper$'s can be compensated by enhancing factors $\ln 1/z$.
Equivalently, we can say that we must give up $\kper$-ordering  and compute
the large $\ln x$-factors associated with the large rapidity gap and {\em any}
value of transverse momenta.

The possible impact of this improved treatment in the small-$x$ region can be
argued from Eqs.~(\ref{fsum}) and (\ref{fgap}). Releasing the
$\kper$-ordering constraint in Eq.~(\ref{fsum}), we miss a $1/n!$ contribution
to Eq.~(\ref{fgap}), thus obtaining
\begin{equation}
\label{flip}
F_{gg} (x,Q^{2};Q^{2}_{0}) \sim \sum_n \frac{1}{n!}
({\mbox {\rm const.}}\; \abar L )^n
\sim\exp \left\{ {\mbox{\rm const.}}\; \abar L \right\} \; ,
\end{equation}
where the constant factor in the exponent is related to the transverse-momentum
integrations. As I shall discuss in the next subsection, the behaviour in
Eq.~(\ref{flip}) is precisely the result of the resummation of the terms
$(\as \ln x)^n$ in perturbation theory.

\subsection{The BFKL equation}

In order to perform this resummation it is convenient to introduce the
unintegrated parton density $\cf(x, \bk;Q_0^2)$ [\ref{BCM}]. It gives the Green
function for the QCD evolution with a fixed total transverse momentum $\bk$
in the final state. The customary Green function $F(x,Q^2;Q_0^2)$ in
Eqs.~(\ref{hatF}) and (\ref{fax}) is then recovered by $\kper$-integration:
\begin{equation}
\label{xfg}
F(x,Q^2;Q_0^2) = \int_{0}^{Q^{2}} d^{2} \bk \;\;{\cal F}(x, \bk;Q_0^2) \;\;.
\end{equation}

In terms of this function the resummation of leading contributions
$(\as \ln x)^n$ is achieved by an integral equation whose kernel describes
parton radiation without $\kper$-ordering constraint:
\begin{equation}
\label{BFKLe}
\cf(x,\bk;Q_0^2)=\pi \,\delta(1-x) \,\delta(\bk^2-Q_0^2) + \abar \int_x^1
\frac{dz}{z} \int d^2\bk' \;K(\bk;\bk') \;\cf(x/z,\bk';Q_0^2) \;.
\end{equation}
The explicit form of the kernel is the following
\begin{equation}
\label{ker}
\int d^2\bk' \;K(\bk;\bk') \;\cf(x,\bk') = \int \frac{d^2\bq}{\pi \bq^2}
\left[ \cf(x,\bk+\bq) - \Theta(\bk^2-\bq^2) \cf(x,\bk) \right] \;
\end{equation}
and was derived by Balitskii, Fadin, Kuraev and Lipatov (BFKL) almost
two decades ago [\ref{BFKL}].

The detailed discussion of Eq.~(\ref{BFKLe}) within the context of the present
paper can be found elsewhere [\ref{GLR},\ref{C},\ref{CFMO}]. Here I limit
myself to recall its main features.

Since the BFKL equation (\ref{BFKLe}) does not enforce $\kper$-ordering, it
is not an evolution equation in $\kper^2$. Its general solution is thus
contamined by higher-twist contributions. Nonetheless, in the large
$\kper$-regime, one can disentangle the leading-twist behaviour in the form
($\cf_N$ are the $N$-moments of $\cf(x)$):
\begin{equation}
\label{fn}
\cf_N(\bk;Q_0^2) = \frac{\ga_N(\as)}{\pi \bk^2} \left( \frac{\bk^2}{Q_0^2}
\right)^{\ga_N(\as)} \left[ 1 + {\cal O}(Q_0^2/\bk^2) \right] \;\;.
\end{equation}
This power behaviour in $\kper$ implies that the resummation of the
$(\as \ln x)$-terms performed by the BFKL equation is consistent with an
effective gluon anomalous dimension $\ga_{gg,N}(\as) \simeq \ga_N(\as)$
[\ref{C},\ref{CFMO},\ref{Jar}]. Its explicit expression is obtained by solving
the following implicit equation
\begin{equation}
\label{andim}
1= \abn \; \chi(\ga_N(\as)) \;\;,
\end{equation}
where the function $\chi(\ga)$ is related to the eigenvalues of the BKFL
kernel and given by ($\psi(z)$ is the Euler $\psi$-function)
\begin{equation}
\label{chi}
\chi(\ga) = 2 \psi(1) - \psi(\ga) -\psi(1-\ga) \;\;.
\end{equation}

The power series solution of Eq.~(\ref{andim}) is ($\zeta(n)$ is the Riemann
zeta function):
\begin{equation}
\label{gaper}
\ga_N(\as) = \sum_{n=1}^\infty c_n \left( \abn \right)^n = \abn + 2 \zeta(3)
\left( \abn \right)^4 + 2 \zeta(5) \left( \abn \right)^6 + {\cal O}\left(
\left( \abn \right)^7 \right) \;\;.
\end{equation}
The one-loop contribution on the r.h.s. of Eq.~(\ref{gaper}) reproduces the
small-$N$ limit (\ref{gammagg}) of the Altarelli-Parisi
splitting function $P^{(1)}$
in Eq.~(\ref{Pz}). The higher orders are the result of the resummation of the
leading logarithmic contributions $(\as^n \ln^{n-1} z)/z$ in $P_{gg}(z,\as)$.

\vspace*{8.5 cm}

\noindent {\bf Fig. 7}: The BFKL characteristic function $\chi(\ga)$ for
$0 < \ga < 1$. $\ga_N$ is the BFKL anomalous dimension.

\vspace*{1 cm}

The characteristic function $\chi(\ga)$ in Eq.~(\ref{chi}) has approximately
a parabolic shape (Fig.~7). As $x \rightarrow 0$, $N$ decreases and reaches a
minimum value $N_{{\rm min}}=4 \abar \ln 2$ at which $\ga_N$ has a branch point
singularity. Therefore the resummation of the singular terms $(\as/N)^n$
builds up a stronger singularity at $N=N_{{\rm min}}$. This singularity,
known as the {\em perturbative QCD} (or BFKL) {\em pomeron}, is responsible
for a very steep behaviour of the QCD evolution at small $x$ (and large
$\kper^2$):
\begin{equation}
\label{pom}
\cf(x,\bk; Q_0^2) \sim \frac{1}{\pi \bk^2} {\sqrt {\frac{\bk^2}{Q_0^2}}}
\,\; x^{-4 \abar \ln 2} \;\;,
\end{equation}
\begin{equation}
\label{Fpom}
F_{gg}(x, Q^2; Q_0^2) = \int_0^{Q^2} d^2\bk \;\cf(x,\bk;Q_0^2) \sim
{\sqrt {\frac{Q^2}{Q_0^2}}} \,\; x^{-4 \abar \ln 2} \;\;.
\end{equation}

I have so far reviewed the main qualitative predictions of perturbative QCD
about the small-$x$ behaviour of the parton densities and, hence, of the
structure function $F_2(x,Q^2)$. The latter is expected to increase faster
that any power of $\ln 1/x$. To leading logarithmic order in $\ln Q^2/Q_0^2$,
the parton densities are driven by the Green function in Eq.~(\ref{Fgg}).
Moreover, at very small values of $x$, the complete (i.e. including subleading
terms in $\ln Q^2/Q_0^2$) resummation of the contributions $(\as \ln x)^n$
gives rise to the power increase in Eq.~(\ref{Fpom}). Note that the transition
from the small-$x$ behaviour in Eq.~(\ref{Fgg}) to that in Eq.~(\ref{Fpom})
is also accompanied by strong scaling violations (compare
$\exp \sqrt {\ln Q^2/Q_0^2}$ with $\sqrt {Q^2/Q_0^2}$).

\subsection{Comment on unitarization}

Before presenting a more detailed quantitative discussion, I should add some
comments on unitarity.

Because of unitarity, the DIS cross section (\ref{dsigma}) cannot increase
indefinitely. At asymptotic energies it must approach a constant (modulo
$\ln S$-corrections) limit as given by the geometrical size of the hadron.
This limit implies the following unitarity bound on the structure function
($R_h$ is the hadron radius)
\begin{equation}
\label{Funi}
\frac{1}{Q^2} \;F_2(x,Q^2) \;\ltap \;\pi \;R_h^2 \;\;.
\end{equation}

If we consider the leading-twist factorization formula (\ref{F2(x}) and the
fact
that gluon evolution dominates at small $x$, from Eq.~(\ref{Funi}) we obtain a
corresponding constraint on the gluon density, namely
\begin{equation}
\label{fguni}
\as \;f_g(x,Q^2) \;\ltap \;\pi \,R_h^2 \,Q^2 \;\;.
\end{equation}
On the other hand, from Eqs.~(\ref{Fgg}) and (\ref{Fpom}), we see that both the
leading-order Altarelli-Parisi equation and the BFKL equation violate the
constraint (\ref{fguni}) at sufficiently small values of $x$. This
contradiction has a simple physical explanation. As soon as the gluon density
becomes very large, the partons tend to overlap and are no longer scattered
incoherently by the hard photon. In this regime we should take into account
parton rescattering and parton recombination: they are responsible for
fulfilling the unitarity bound in Eq.~(\ref{Funi}).

{}From a formal viewpoint, this means we have to include higher-twist
contributions. Parton recombination is a process of order $1/(R_h^2 Q^2)$ and
the probability of recombination of two gluons is approximately given by $\as
(f_g(x,Q^2))^2$. Therefore this process produces an higher-twist correction of
relative order  $\as f_g(x,Q^2)/(R_h^2 Q^2)$ on the r.h.s. of
Eq.~(\ref{F2(x}).
As soon as this corrections approaches unity, higher-twist terms cannot be
neglected in the factorization formula (\ref{F2(x}). We have thus recovered the
constraint (\ref{fguni}), which has to be correctly interpreted as a limit on
the validity of the leading-twist approximation.

This argument shows that the perturbative QCD approach to small-$x$ physics is
(at least) self-consistent as regards to unitarity. The factorization formula
(\ref{F2(x}) has to be used in a proper way by including higher-twist
corrections whenever they are (estimated to be) large.

In spite of many efforts and partial results [\ref{uni}], a general theory of
rescattering has not yet been set up. Reasonable physical models are however
available [\ref{GLR}]. Recent numerical estimates [\ref{Ask}] based on these
models lead to the conclusion that rescattering effects give a small
contribution (at most $10\%$) to $F_2$ in the HERA kinematic range. On the
other side, at present, no evident signal of unitarity corrections
(i.e. flattening of $F_2$ at small $x$)
emerges from the HERA data. This is not surprising
since, even for values of $Q^2$ as small as 1 GeV${}^2$, the bound
(\ref{Funi}) gives $F_2 \,\ltap \,25 \;(\pi R_h^2 \simeq 1$ fm${}^2$).
Because of these reasons, in the following I shall not discuss any longer
the issue of unitarity.

\section{NLO parton densities}

In order to perform a quantitative comparison between HERA data and
perturbative QCD predictions, I shall consider some of next-to-leading
order (NLO) parton densities that were available before HERA started operating.
These parton densities are obtained by evolving (as in Eq.~(\ref{fax})) some
input distributions according to the generalized Altarelli-Parisi equation
(\ref{Fab}) in NLO, i.e. including the full one- and two-loop [\ref{2loop}]
splitting functions (anomalous dimensions) in Eq.~(\ref{Pz})
(Eq.~(\ref{gammaN})).
The input parton densities at some scale $Q_0^2$  are fixed by fitting
pre-HERA data on muon and neutrino DIS, prompt-photon production and Drell-Yan
type processes in hadron-hadron collisions. The ensuing predictions have a
high (better than $10\%$) overall accuracy, including the theoretical
uncertainties due to higher-twist effects and perturbative corrections beyond
two-loop order. For this reason, they provide a good starting point for
detailed
quantitative investigations of the small-$x$ region.

Actually, since pre-HERA data concern the high-$x$ region, the input gluon
and sea-quark distributions $f_g$ and $f_{{\rm sea}}$ are essentially
unconstrained for values of $x$ smaller than $10^{-2}$. The representative
sets of NLO parton densities listed below precisely differ on the assumptions
about the small-$x$ behaviour of the input densities.

The GRV parton distributions [\ref{GRV}] are obtained by starting the
perturbative QCD evolution at a very low input scale, $Q_0^2=0.3$ GeV${}^2$.
At this scale the input distributions are chosen to vanish for
$x \rightarrow 0$:
$f_g(x) \sim x^2, \; f_{\rm sea}(x) \sim x^{0.7}$.
Because of this `valence-like'
behaviour of the input, any steeper behaviour at high $Q^2$ is entirely
generated by perturbative QCD evolution.

The input scale used in the MRS analysis [\ref{MRS}] is $Q_0^2=4$ GeV${}^2$.
They provide two representative sets of parton densities. Set D'${}_0$ is
obtained using input densities which are flat at small $x$:
$f_g(x) \sim f_{\rm sea}(x) \sim $ const. . Set D'${}_{-}$ instead uses
very steep input distributions:
$f_g(x) \sim f_{\rm sea}(x) \sim x^{-0.5}$. Roughly speaking, the flat input
corresponds to the so-called `critical-pomeron' (with intercept equal to unity)
behaviour whereas the steep input is reminiscent of the BFKL pomeron.

As in the case of the MRS analysis, the CTEQ Collaboration [\ref{CTEQ}] starts
the perturbative evolution at the input scale $Q_0^2= 4$ GeV${}^2$.
Among the different sets
of parton distributions they consider, set CTEQ1MS is that which has the
steepest input densities, namely
$f_g(x) \sim x^{-0.38} (\ln 1/x)^{0.09}, \; f_{\rm sea}(x) \sim x^{-0.27}$.
This small-$x$ behaviour is just the result of a certain parametrization of
the input and is not related to any `hybrid pomeron'.

Figures 5 and 8 show the comparison between the 1992
data\footnote{The comparison with the 1993 preliminary data of the
ZEUS Collaboration, presented at this meeting [\ref{Newton}], does not
substantially modify the discussion that follows.}
of the H1 and ZEUS Collabo\-rations
and the structure function
$F_2$ as obtained from different sets of NLO parton densities. We see that the
expectations from the GRV and MRSD'$_{-}$ sets seem favoured by the data.

\vspace*{17 cm}

\noindent {\bf Fig. 8}: Comparison of the measured  $F_2(x,Q^2)$ (ZEUS data)
with the expectations from NLO parton densities.
The overall normalization uncertainty of $7\%$ is not included
in the error bars.

\vspace*{1 cm}

This observation, rather than being the main conclusion of the present
contribution, gives me the opportunity for addressing some points which require
further investigations and that I consider relevant for the understanding of
small-$x$ physics.

The GRV parton distributions on one side and the MRSD'$_0$, MRSD'$_{-}$,
CTEQ1MS
distributions on the other side
represent, in a sense, two very different QCD expectations.

In the GRV framework, due to the valence-like behaviour of the input densities,
the fast increase of $F_2$ at small-$x$ and high $Q^2$ is entirely generated
by perturbative QCD evolution as in Eq.~(\ref{Fgg}). More precisely, according
to Eq.~(\ref{Fgg}), a structure function as steep as that reported in Figs.~5
and 8 can be obtained only by means of a very long $Q^2$-evolution. In the GRV
case this is achieved by using a very low input scale. The weak (or, perhaps,
amazing) point is that no higher-twist contribution is included (present) even
at an input scale as low as $Q_0^2=0.3$ GeV${}^2$. Moreover, no small-$x$
resummation (either from the BFKL equation or from higher orders in the
Altarelli-Parisi equation) is considered.

In the case of the MRS and CTEQ parton distributions the starting scale for
the NLO QCD evolution is reasonably large, namely $Q_0^2=4$ GeV${}^2$. The
different expectations for $F_2$ at small $x$ are entirely due to the
different behaviour of the input densities. The MRSD'$_{-}$ set, which is
favoured by the data, has the steepest input densities. According to a
widespread wisdom, the MRSD'$_{-}$ framework is motivated by the BFKL equation.
If one considers the BFKL pomeron intercept $N_{min}= 4\abar \ln 2 \simeq
2.65 \;\as$ and a `reasonable' value for $\as$, say $\as \simeq 0.2$, one gets
$N_{min} \simeq 0.5$: $x^{-0.5}$ is indeed the small-$x$ behaviour of the input
parton densities in the MRSD'$_{-}$ set. However, the {\em perturbative} (BFKL)
pomeron is used here as {\em non-perturbative} input. On the contrary, the
perturbative QCD evolution is performed with the NLO Altarelli-Parisi equation.
In particular, since the input densities have an $x$-shape which is steeper
than the one generated by the NLO QCD evolution, it follows (see the comment
before Eq.~(\ref{hatF})) that the small-$x$
behaviour of $F_2$ is dominated by that of the non-perturbative input. I find
this situation a little uncomfortable, although consistent at present. As  a
matter of fact, there are theoretical and phenomenological problems which are
still to be solved.

As regards to the theory, the BFKL equation resums the leading terms
$(\as \ln x)^n$ in QCD perturbation theory but it is {\em not} an evolution
equation in $\kper^2$. Its matching with the Altarelli-Parisi equation at
larger
values of $x$ (or $Q^2$) and its extension (if any) to subleading orders
in $\ln x$
requires the inclusion of the running coupling $\as(\kper^2)$. Due to
the lack of $\kper$-ordering, this leads to infrared instabilities coming
from the Landau pole of $\as(\kper^2)$ in the low-$\kper$ (non-perturbative)
region.

As regards to phenomenology, in any perturbative expansion the leading term
predicts just the order of magnitude of a given quantity. The accuracy of the
prediction is instead controlled by the size of the higher-order corrections.

The inclusion of subleading terms at small $x$ is therefore necessary for
precise quantitative studies.

\section{Beyond leading order at small $x$}
%$\bf x$}

A possible approach [\ref{CH},\ref{QAD}] for improving and controlling
the accuracy of the
perturbative QCD predictions at small $x$ (and large $Q^2$) consists in
combining small-$x$ resummation with leading-twist factorization.

The generalized Altarelli-Parisi equation (\ref{Fab}) for the
evolution of the parton densities to leading-twist order
systematically takes into account
QCD corrections via the perturbative expansion (\ref{Pz})
of the splitting functions
$P_{ab}(z,\as)$. However, as discussed in Sects.~4.2 and 4.3, higher-loop
contributions to $P_{ab}$ are logarithmically enhanced at small $x$. More
precisely, in the small-$x$ limit, the $n$-loop order splitting function
$P_{ab}^{(n)}$ behaves as
\begin{equation}
P_{ab}^{(n)}(x) \sim \frac{1}{x} \left[ \ln^{n-1} x + {\cal O}(\ln
^{n-2}x) \right] \;\;.
\label{Pabn}
\end{equation}
Correspondingly, the anomalous dimensions $\ga_N$ in Eq.~(\ref{gammaN}) have
singularities for $N \rightarrow 0$ in the form:
\begin{equation}
\gamma_{ab,N} {(\as)} = \sum^{\infty}_{n=1} \left[
\left(\frac{\as}{N}\right)^{n} A^{(n)}_{ab} + \as
\left(\frac{\as}{N}\right)^{n} B^{(n)}_{ab} + \dots \right] \;,
\label{gamab}
\end{equation}

These singularities may spoil the convergence of the perturbative
expansion at small $x$. Nonetheless one can consider an improved
perturbative expansion
obtained by resumming the {\em leading}
$(A_{ab}^{(n)})$, {\em next-to-leading}
$(B_{ab}^{(n)})$,
%$\dots$
etc.,
coefficients in the small-$x$ regime. Once these coefficients are known, they
can unambiguously be supplemented with non-logarithmic (finite-$x$)
contributions exactly calculable to any fixed-order in perturbation theory.
Moreover, the effects of the running coupling can be consistently included
according to Eq.~(\ref{FNQ}),
thus avoiding the infrared instabilities encountered
in phenomenological attempts [\ref{Ask},\ref{qzero}] to extend the
BFKL equation beyond
leading order. Obviously, in this approach, the absolute behaviour of the
structure function $F_2$ at small $x$ is still affected by that of the input
parton densities. However, since logarithmic scaling violations are
systematically under control, our confidence in making quantitative predictions
at high $Q^2$ is enhanced.

The present status of small-$x$ resummation is the following. The gluon
anomalous dimensions $\ga_{gg,N}$ contain leading singularities of the type
$(\as/N)^n$. These contributions are given by the BFKL anomalous dimensions in
Eq.~(\ref{gaper}). The next-to-leading corrections $\as(\as/N)^n$ in the gluon
sector are still unknown. Beyond leading accuracy, the quark sector has to
be considered on an equal footing with the gluon sector. The next-to-leading
corrections to the quark anomalous dimensions have been computed recently
[\ref{QAD}]. The first perturbative terms are explicitly given by ($T_R=1/2$
in QCD)
\begin{equation}
\label{qms}
\ga_{qg,N}(\as) = \frac{\as}{2\pi} T_R \;\frac{2}{3}
\left\{ 1 + \frac{5}{3} \abn +
\frac{14}{9} \left(\abn \right)^2 +
+ {\cal O}\!\left( \left(\abn \right)^3 \right) \right\} \;\;,
\end{equation}
%\begin{eqnarray}
%\label{qms}
%\ga_{qg,N}(\as) &=& \frac{\as}{2\pi} T_R \;\frac{2}{3}
%\left\{ 1 + \frac{5}{3} \abn +
%\frac{14}{9} \left(\abn \right)^2 +
%\left[\frac{82}{81}+ 2 \, \zeta (3)    \right] \,
% \left(\abn \right)^3
%+ \left[ {122 \over 243}
%\right.
%\right.
%\nonumber\\
%&+& \left.
%\left.
% {25 \over 6} \, \zeta (3) \right] \,
% \left(\abn \right)^4
%+ \left[ {146 \over 729}+{14 \over 3}\,\zeta (3)+2\,\zeta (5) \right] \,
% \left(\abn \right)^5
%+ {\cal O}\!\left( \left(\abn \right)^6 \right) \right\} \nonumber \\
%&\simeq& \frac{\as}{2\pi} T_R \;\frac{2}{3} \left\{ 1 + 1.67 \abn +
%1.56 \left(\abn \right)^2 +
%3.42 \left(\abn \right)^3 \right.
%\nonumber\\
%&+& \left.
%5.51 \left(\abn \right)^4 +
%7.88 \left(\abn \right)^5 +
%{\cal O}\!\left( \left(\abn \right)^6 \right) \right\} \;\;.
%\end{eqnarray}
and higher-loop contributions can be found in Ref.~[\ref{CH}].

A numerical analysis of the effect of next-to-leading corrections at small $x$
has been presented recently by Ellis, Kunszt and Levin [\ref{EKL}]. They
consider both the flat (D'$_0$) and steep (D'$_{-}$) input distributions used
by MRS and, besides carrying out a full two-loop evolution  (as in the MRS
analysis), they include also the ${\cal O}((\as/N)^4)$-term of the gluon
anomalous dimensions (\ref{gaper}) and the ${\cal O}((\as/N)^3)$-term of the
quark anomalous dimensions (\ref{qms}). Their results on the structure function
$F_2$ are reported in Fig.~9. We can see that, for $x \,\ltap \,10^{-2}$,
the effect of the higher-order
corrections strongly depends on the $x$-shape of the
input distributions. The $Q^2$-evolution of the steep input is marginally
affected by the inclusion of higher-loop contributions\footnote{A similar
perturbative stability is likely to occur in the GRV framework because
of the very low input scale $Q_0$ and, hence, of the very long evolution in
$Q^2$.}. On the contrary, the $Q^2$-evolution of the flat input is
pertubatively
unstable\footnote{Since in DIS the virtual photon couples directly to quarks
(and not gluons), the quark anomalous dimensions (\ref{qms}) (despite being
formally subleading with respect to the gluon anomalous dimensions in
Eq.~(\ref{gaper})) are largely responsible for the perturbative
instability in the HERA region.}.

A first conclusion is that, {\em if} the input distribution is very steep,
higher-order perturbative corrections are probably negligible, at least in the
small-$x$ region which is being investigated at HERA. On the other side, this
does not mean that a flat input has to be excluded due to its perturbative
instability. In this case one should rather include higher-loop
corrections at small $x$
and, possibly, recover reliable predictions after their all-order
resummation. The difference between these two alternatives (steep input +
fixed-order and flat input + resummation) is not just a matter of nominalism.
Although the respective predictions for $F_2$ at a certain value of $Q^2$ can
be very similar, the scaling violations (i.e. the $Q^2$-dependence) are
stronger in the second case. Further studies are necessary for quantifying
precisely the phenomenological consequences of this discussion for the HERA
kinematic region, as well as, at future hadron colliders.

\pagebreak

\vspace*{8.5 cm}

\noindent {\bf Fig. 9}: Results of the QCD evolution of the steep (upper
curves) and flat (lower curves) input distributions. The effect of including
higher-order terms (NLL) is compared with the full one-loop (L) and
two-loop (NL) evolution.

\section{Conclusions}

The structure function $F_2(x,Q^2)$ measured at HERA increases very steeply
for $x\, \ltap \,10^{-2}$.
A similar rise with the energy is not observed for total cross sections in soft
hadronic processes. Therefore the structure function data at low-$x$ cannot be
explained without hard QCD contributions.

Perturbative QCD qualitatively accounts for the steep increase of $F_2$. As
regards to the sets of NLO parton densities available before HERA, the GRV and
MRSD'$_{-}$ parton densities are favoured by the HERA data. Both sets of parton
densities are obtained by perturbative QCD evolution in two-loop order. In
the GRV framework `valence-like' input densities at a very low input scale
undergo a very long $Q^2$-evolution, thus acquiring the steep behaviour given
by the Altarelli-Parisi equation (\ref{Fgg}). In the MRSD'$_{-}$ framework,
input distributions as steep as $x^{-0.5}$ (inspired by the BFKL equation) are
considered at a much higher input scale. The low-$x$ behaviour of $F_2$ at high
$Q^2$ is thus dominated by that of the input densities\footnote{The small-$x$
behaviour of the input densities can be `tuned' to describing better $F_2$.
Recent fits to the 1992 HERA data, carried out by MRS and the CTEQ
Collaboration, give the behaviour $x^{-0.3}$ [\ref{New}].}.

Estimating the present accuracy of the perturbative QCD predictions requires
the evaluation of higher-order contributions and more detailed quantitative
studies. For instance, from the discussion in Sect.~6, one can argue that
flat (or almost flat, as suggested by the soft-pomeron behaviour [\ref{DOLA}])
input densities at a scale of the order of 1 GeV$^2$, combined with a
dynamically generated perturbative pomeron (namely, all-order resummation
in the Altarelli-Parisi splitting functions), may lead to structure functions
consistent with HERA data.

On the theoretical side, more calculations on next-to-leading terms at small
$x$
are necessary. On the phenomenological side, I consider relevant the analysis
of the $Q^2$-dependence and of the scaling violations. Studies on small-$x$
effects in the structure of the hadronic final states are also important
[\ref{CFMO},\ref{GDTK},\ref{Webber},\ref{et}].
More (and more accurate) experimental information
will certainly be available in the very near future.

\vskip 1 true cm

\noindent {\bf Acknowledgements} -- I wish to thank Giorgio Bellettini and
Mario Greco for the pleasant and stimulating atmosphere at this Conference.

\vskip 1 true cm

{\large \bf References}
\begin{enumerate}
%\normalsize

\item \label{Pope}
      B.\ Pope, these proceedings; J.\ Smith, these proceedings.

\item \label{H1}
      H1 Coll., I.\ Abt et al., \np{407}{515}{93}.

\item \label{ZEUS}
      ZEUS Coll., M.\ Derrick et al., \pl{316}{412}{93}.

\item \label{Newton}
      D.\ Newton, these proceedings.

\item \label{Aachen}
      M.\ Virchaux, in Proc. of the Aachen Conf. {\it QCD - 20 years later},
      eds. P.M.\ Zerwas and H.A.\ Kastrup (World Scientific, Singapore, 1993),
      pag.~205 and references therein.

\item \label{PDG}
      Particle Data Group, K.\ Hikasa et al., Review of particle properties,
      Phys. Rev. D45, No.~11 (1992).

\item \label{DOLA}
      A.\ Donnachie and P.V.\ Landshoff, \np{231}{189}{83}, \pl{296}{227}{92}.

\item \label{phot}
      ZEUS Coll., M.\ Derrick et al., \pl{293}{465}{92};
      H1 Coll.,  T.\ Ahmed et al., \pl{299}{374}{93}.

\item \label{DOLA1}
      A.\ Donnachie and P.V.\ Landshoff, \np{244}{322}{84}.

\item \label{ALLM}
      H.\ Abramowicz, E.M.\ Levin, A.\ Levy and U.\ Maor, \pl{269}{465}{91}.

\item \label{DOLA2}
      A.\ Donnachie and P.V.\ Landshoff, Cambridge preprint DAMTP 93-23.

\item \label{Alt}
      G.\ Altarelli, \prep{81}{1}{82} and references therein.

\item \label{CCH}
      S.\ Catani, M.\ Ciafaloni and F.\ Hautmann, Phys. Lett. B242
      (1990) 97, Nucl. Phys. B366 (1991) 135.

\item \label{Neerven}
      E.B.\ Zijlstra and W.L.\ van Neerven, \np{383}{525}{92}.

\item \label{CH}
      S.\ Catani and F.\ Hautmann, \cav{94/01}.

\item \label{BCM}
      A. Bassetto, M. Cia\-fa\-lo\-ni and G. Mar\-che\-si\-ni,
      Phys. Rep. 100 (1983) 201.

\item \label{BFKL}
      L.N.\ Lipatov, Sov. J. Nucl. Phys. 23 (1976) 338; E.A.\ Kuraev,
      L.N.\ Lipatov and V.S.\ Fadin, Sov. Phys. JETP  45 (1977) 199; Ya.\
      Balitskii and L.N.\ Lipatov, Sov. J. Nucl. Phys. 28 (1978) 822.

\item \label{GLR}
      L.V.\ Gribov, E.M.\ Levin and M.G.\ Ryskin, \prep{100}{1}{83}.
%      E.M. Levin and M.G. Ryskin, Phys. Rep. 189 (1990) 268.

\item \label{C}
      M.\ Cia\-fa\-lo\-ni, \np{296}{249}{87}.

\item \label{CFMO}
      S.\ Catani, F.\ Fiorani and G.\ Marchesini, \pl{234}{339}{90},
      \np{336}{18}{90}; S.\ Catani, F.\ Fiorani, G.\ Marchesini and G.\ Oriani,
      \np{361}{645}{91}.

\item \label{Jar}
       T. Jaroszewicz, Phys. Lett. B116 (1982) 291.

\item \label{uni}
      A.H. Mueller and J. Qiu, Nucl. Phys. B268 (1986) 427;
      A.H. Mueller, Nucl. Phys. B335 (1990) 115;
      J.\ Bartels, preprint DESY-91-074, \pl{298}{204}{93}, \zp{60}{471}{93};
%preprint ESY-93-028;
      E.M.\ Levin, M.G.\ Ryskin and A.G.\ Shuvaev,
      Nucl. Phys. B387 (1992) 589;
      E.\ Laenen, E.M.\ Levin and A.G.\ Shuvaev, preprint
      Fermilab-PUB-93/243-T.

\item \label{Ask}
%\item \label{Dur}
       A. J. Askew, J. Kwiecinski, A. D. Martin and P. J. Sutton,
       Phys. Rev. D 47 (1993) 3775, \pr{49}{4402}{94}.
%       preprint DTP-93-28.

\item \label{2loop}
      E.G.\ Floratos, D.A.\ Ross and C.T.\ Sachrajda, \np{129}{66}{77}
      (E \np{139}{545}{78}), \np{152}{493}{79}; A.\ Gonzalez-Arroyo, C.\ Lopez
      and F.J.\ Yndurain, \np{153}{161}{79}; A.\ Gonzalez-Arroyo and C.\ Lopez,
      \np{166}{429}{80};
      G.\ Curci, W.\ Furmanski and R.\ Petronzio,  \np{175}{27}{80};
      W.\ Furmanski and R.\ Petronzio, \pl{97}{437}{80};
      E.G.\ Floratos, P.\ Lacaze and C.\ Kounnas,
      \pl{98}{89}{81}, 225.

\item \label{GRV}
      M.\ Gl\"{u}ck, E.\ Reya and A.\ Vogt, \zp{53}{127}{92},
\pl{306}{391}{93}.

\item \label{MRS}
      A.D.\ Martin, R.G.\ Roberts and W.J.\ Stirling, \pl{306}{145}{93}.

\item \label{CTEQ}
      CTEQ Coll., J.\ Botts et al., \pl{304}{159}{93}.

\item \label{QAD}
      S.\ Catani and F.\ Hautmann, \pl{315}{157}{93}.

\item \label{qzero}
      J.\ Collins and J.\ Kwiecinski, \np{316}{307}{89};
      J.\ Kwiecinski, A.D.\ Martin and P.J.\ Sutton, \pr{44}{2640}{91}.

\item \label{EKL}
      R.K. Ellis, Z. Kunszt and E. M. Levin, preprint Fermilab-PUB-93/350-T.

\item \label{New}
      A.D.\ Martin, R.G.\ Roberts and W.J.\ Stirling, Durham preprint
      DTP/93/86; CTEQ Coll., J.\ Botts et al., CTEQ2 sets of parton
      distributions.

\item \label{GDTK}
      L.V.\ Gribov, Yu.L.\ Dokshitzer, S.I.\ Troyan and V.A.\ Khoze, Sov. Phys.
      JETP 67 (1988) 1303.

\item \label{Webber}
      G.\ Marchesini and B.R. Webber, \np{349}{617}{91}, \np{386}{215}{92};
      B.R.\ Webber, in Proceedings of the HERA Workshop,
      eds. W.\ Buchm\"{u}ller and G.\ Ingelman (DESY Hamburg, 1991), pag.~285.

\item \label{et}
      J.\ Kwiecinski, A.D.\ Martin, P.J.\ Sutton and K.\ Golec-Biernat,
      Durham preprint
\linebreak
      DTP/94/08.

\end{enumerate}

\end{document}